
\magnification\magstep1
\hsize5.45truein
\hoffset.8truein
\baselineskip23truept
\topskip10truept
{
\nopagenumbers
\hfill{IP-ASTP-06-92}
\vfill
\centerline{\bf A Hydrodynamical Analysis of the Burning of a Neutron Star}
\vskip.5in

\centerline{H. T. Cho$^1$}
\vskip.25truecm
\centerline{Department of Physics}
\centerline{Tamkang University}
\centerline{Tamsui, Taipei, Taiwan 25137}
\vfill
\centerline{K. -W. Ng$^2$}
\vskip.25truecm
\centerline{and}
\vskip.25truecm
\centerline{A. D. Speliotopoulos$^3$}
\footnote{}
{$^1$ PHCHO@TWNAS886.\hfill\break\indent $^2$
PHKWNG@TWNAS886\hfill\break\indent $^3$ PHADS@TWNAS886}

\vskip.5cm

\centerline{\it Institute of Physics}
\centerline{\it Academia Sinica}
\centerline{\it Nankang, Taipei, Taiwan 11529}

\vfill

\centerline{\bf Abstract}

\vskip.25in

{\leftskip2in
 \rightskip2in
 \noindent The burning of a neutron star by strange matter is
 analyzed using relativistic combustion theory with a planar
 geometry. It is shown that such burning is probably neither slow
 combustion nor simple detonation. Fast combustion without detonation
 is possible under certain circumstances, but would involve very
 efficient heat transfer mechanisms. It is found, however, that the
 burning is most likely absolutely unstable with no well defined burn front.
}
\vskip1truecm
\vfill
\supereject
}
\pageno=2
\noindent{\bf \S 1. Introduction.}

There have recently been a great deal of interest $[1]-[7]$ in the conversion
of neutron stars into ``strange'' stars. As Witten $[8]$ first proposed,
nuclear matter may not be the most stable form of matter. Rather,
``strange'' matter, matter consisting of equal numbers of up, down and
strange quarks, is. The average energy per nucleon was shown by Fahri and
Jaffe $[9]$ to be lower for this ``strange'' matter than for
nuclear matter. It is believed that if a lump of strange matter comes
in contact with a neutron star by whatever means possible, the
neutron star will quickly be converted into a strange star. The
question is how fast and under precisely what conditions this
conversion will take place.

There are currently two methods being used to analyze this conversion
process. The first is due to Olinto $[2]$ who uses a non-relativistic
diffusion model. As such, theirs is a slow combustion
model, with the burn front propagating at a speed of approximately $10$ m/sec.
This is determined primarily by the rate at which
one of the down quarks inside the neutron is converted through a weak
decay to a strange quark: $d+u\to s+u$.
The second method was introduced by Horvath and Benvenuto
$[6]$ who models the conversion as a detonation. Their conversion
rate is several orders of magnitude faster than that
predicted by Olinto. We, on the other hand, shall not {\it a priori\/} assume
that the conversion process is due to either a slow combustion or a
detonation. In fact, one of the purposes of our analysis is to
establish under what conditions one will have a slow burning of the
star or a detonation. Rather, we shall use the standard theory of relativistic
combustion to analyze the conversion of the star which depends solely on
the equations of state in the two media and the conservation of
energy-momentum and baryon number across the combustion front. It
is a kinematic analysis and has the advantage of not only being independent of
the details of how the neutrons are absorbed into the strange matter,
which at this point is not well understood, but also independent of
any assumptions as to the rate at which the conversion will take
place. Since the only equation of state for the strange
matter is the MIT bag model, any analysis of the conversion of the
neutron star into a strange star will be dependent, to a certain
extent, on the value of the bag
constant, which is not known to any great degree of accuracy. We therefore
calculated the velocity of the conversion front for a
wide range of values of the bag constant and the density of the
neutron star at various temperatures.

\noindent{\bf\S 2. Methodology.}

We shall effectively be working in one spatial dimension, with the
strange matter
and the nuclear matter being seperated by a well defined planar
combustion wavefront $[10]$. It has been shown that this wavefront
is quite abrupt $[6]$, $[3]$, meaning that the thickness $\delta$ of
the wavefront is much smaller than the size of the neutron star. We
shall work in the frame which is at rest with the
combustion wavefront. The energy-momentum tensor for the whole system is
$$
T_{\mu\nu} = (\epsilon+p) u_\mu u_\nu-pg_{\mu\nu}
$$
where $u_\mu$ is the covariant fluid velocity, $\epsilon$ is the
energy density and $p$ is the pressure. The baryonic number density is
$N_\mu = nu_\mu$. We will be using the ideal gas
approximation for the quarks in the strange matter with all the quark
masses $m_q=0$ and the strong coupling constant $\alpha_s
=0$. In this zeroth order approximation the fraction of
electrons in the strange quark matter $Y_e\equiv n_e/n_q=0$ and the energy
loss by neutrino emission from the quark matter vanishes $[11]$.
Even though the process of absorbing neutrons is very exothermic, with each
absorbed neutron releasing $\sim$10 MeV of energy $[9]$, we shall
consider that the reaction, as a whole, conserves energy with the
excess energy released in the absorption process going into heating
up the star as a whole. In more realistic processes this
approximation need not hold, and we shall comment on these situations
in $\bf \S 3$. Then $T_{\mu\nu}$ is conserved across
the combustion front, and it can be shown that $[10]$,
$$
v_q^2 =
{(p_q-p_n)(\epsilon_n+p_q)\over(\epsilon_q-\epsilon_n)(\epsilon_q+p_n)}
\qquad,\qquad
v_n^2 =
{(p_n-p_q)(\epsilon_q+p_n)\over(\epsilon_n-\epsilon_q)(\epsilon_n+p_q)}
\eqno(1)
$$
where $v_q$ and $v_n$ are the velocities relative
to the burn front of the quarks and the neutrons
in the strange matter and in the neutron star respectively. $v_n$ is also
the velocity of the front relative to the star.
(Variables with subscript $q$ ($n$) will denote quantities measured in the
stran
   ge
(nuclear) matter.) Next, baryon number must be conserved
across the wavefront, from which we obtain
$$
n_q^2 = n_n^2{(\epsilon_q + p_q)(\epsilon_q+p_n)\over
        (\epsilon_n + p_n)(\epsilon_n+p_q)}.
$$

For the strange matter, we shall assume a zero strange quark mass and
take the MIT bag model equation of state:
$$
\eqalign{
e_q=&{19\over 12}\pi^2 T^4 + {9\over 2}T^2\mu^2 + {9\over{4\pi^2}}\mu^4 + B\cr
p_q=&{19\over 36}\pi^2 T^4 + {3\over 2}T^2\mu^2 + {3\over{4\pi^2}}\mu^4 - B\cr
n_q=&{T^2\mu + {1\over \pi^2}\mu^3}
}
\eqno(2)
$$
where $T$, $\mu$, and $B$ are the temperature, chemical potential of
each quark, and bag constant respectively.
For the nuclear matter there are various
equations of state that we can choose. The two that we shall work with are
the zero-temperature Bethe-Johnson (BJ) equation of state $[12]$\null
$$
\eqalign{
\epsilon_n=& (236 n_n^{1.54}+m_n)n_n \>\>\hbox{MeV fm$^{-3}$},
        \cr
p_n = &364n_n^{2.54}\>\>\hbox{MeV fm$^{-3}$}
}
\eqno(3)
$$
where $m_n$ is the mass of the neutron in MeV,
and the zero-temperature ideal Fermi-Dirac (FD) neutron gas $[13]$.
The velocities $v_q$ and $v_n$ may now be determined for
any given values of $B$, $n_n$ and $T$. Here we require $0.3\>\>{\rm fm^{-4}}
<B$ so that nuclei with high atomic numbers would be stable against decay
into non-strange quark matter $[9]$.
In addition, requiring the star to be hydrodynamically stable against small
perturbations limits $n_n<1.5\>\>{\rm fm^{-3}}$ for the hard BJ equation of
state, and $n_n<5\>\>{\rm fm^{-3}}$
for the soft FD equation of state $[12]$.

At this point we need to introduce some terminology $[10]$.
Let $v^s_q$ ($v^s_n$) be the speed of sound in the strange (nuclear)
matter. $v^s_q={1\over \sqrt 3}$ while
$$
v_n^s=\bigl [{{n^{1.54}}\over {1.01+0.648n^{1.54}}} \bigr ] ^{1\over 2}
$$
is the speed of sound for the BJ equation of state.
When $v_q<v^s_q$ and $v_n<v^s_n$, the burning is called a
{\it deflagration} or a slow combustion. When $v_q\le v^s_q$ and $v_n>v^s_n$,
it is called a {\it detonation}. When $v_q>v^s_q$ and $v_n<v^s_n$, the burning
is {\it absolutely unstable}, meaning that in the presence of any small
perturbation the wavefront will no longer remain as a well defined
plane and the model itself fails. When $v_q>v^s_q$ and $v_n>v^s_n$,
the burning is a fast combustion without detonation which may involve
either very efficient heat transfer in the unburnt gas, or reactions
which are initially exothermic but are endothermic in their final
stages.

\noindent{\bf\S 3. Results.}

There are a few constraints that must be imposed. First, the
velocities $v_q$ and $v_n$ must be real and less than the speed of
light. Next, for the strange matter to consume the neutron star, the
energy per ``baryon'' in the strange matter must be less than the
energy per baryon in the neutron star. The demarcation lines are the
$\epsilon_q/n_q = \epsilon_n/n_n$ lines and along these lines the bag
constant $B$ may then be solved in terms of $n_n$ for any
given $T$. Plots of the lines
of equal energy per baryon in the $B$-$n_n$ parameter space are shown
with the solid curves in Fig. 1-4 for various choices of $T$.
In each case, the thick solid curve also corresponds to the equal
energy line: $\epsilon_q = \epsilon_n$. Along this line the
velocities are infinite. Consequently, the regime of physically acceptable
velocities for which there is a conversion of the neutron star into a
strange star is confined to be below the lowest $\epsilon_q/n_q =
\epsilon_n/n_n$, to the left of the
vertical line corresponding to the maximum allowed value of $n_n$,
and above the horizontal line corresponding to the minimum allowed
value of $B$.
Notice also that from Eq. $(1)$ the size of $v_q$ and $v_n$ are
determined to a
large extent by the pressure difference between the nuclear
matter and the strange matter. It is only
when the $p_q=p_n$ line lies below the $\epsilon_q=\epsilon_n$ line that
there will be choices of $B$, $n_n$, and $T$ for which $v_q$ or
$v_n$ will be substantially less than the speed of light.

The parameter spaces obtained using the FD equation of state are shown in
Figs. 1-2.
Unlike the BJ equation of state the region of the parameter space for
which burning will occur is bounded to the right by the equal energy
line. When $T=0$ this region is contained within a very small triangular
region on the right of the graph. As the temperature increases,
the $\epsilon_q/n_q = \epsilon_n/n_n$ lines collapse towards the
$n_n$ axis and we find that at $T=60$ MeV this region disappears
altogether.
We thus conclude that the ideal fermi neutron gas is in general too ``soft''
to allow the neutron star to burn.

Figs. 3-4 show the graphs of the parameter space using the BJ
equation of state. Unlike the FD equation of state, there does exist
choice of $B$ and $n_n$ such that burning may occur and we have
included in these graphs contour lines of constant $v_q$ (dashed
lines) and constant $v_n$ (dotted lines). When $T=0$ we found that
there was no choices of $B$ and $n_n$ for which $v_q<0.8$ so that
$v_q>v_q^s$ throughout the parameter space. Consequently, the starred line,
representing the $v_n=v_n^s$ contour, splits the physically allowed
parameter space only into two regions. Below this line $v_q>v_q^s$ while
$v_n<v_n^s$. Consequently, in this region the burning is unstable.
Above this line $v_q>v_q^s$ and $v_n>v_n^s$. This is the exotic
region in which the burning can only occur due to some exotic heat
transfer mechanisms.

As the temperature increases, the $\epsilon_q/n_q =
\epsilon_n/n_n$ lines once again collapse towards the $n_n$ axis, but
because the physical region is not bounded to the right by the equal
energy line, there will still be a region in which the burning will
take place. When $T=60$ MeV we find that once again we have an exotic
region and an unstable region in the parameter space, but not
we also have a small region lying to the left of the $v_q=0.6$ contour in
which the star will undergo a ``slow'' combustion. (As the
temperature is increased further, this region eventually also
disappears.) Even in this slow combustion region $v_n$ is still typically
fractions of the speed of light, however. More importantly, we find
that in this region $v_n<v_q$ and consequently the burning is
unstable $[10]$. Moreover, by incorporating gravitational effects
inside the neutron star and also a surface tension term ascribed to
the burn front, Horvath and Benvenuto $[6]$ has shown that the slow
combustion is unstable and is likely to become a detonation. In both
Figs. 3 and 4 we have included the large dotted line corresponding to
$\epsilon_n/n_n-\epsilon_q/n_q\approx 10$ MeV. The most probable
velocities for the burning of the neutron star should occur along
this line.

We find that the conversion of a neutron star into a strange star is
{\it never\/} due to a detonation, although, except for some judicious
choices of parameters, it will occur extremely rapidly. If the
fast combustion is stable, then some presently unknown exotic mechanism
must be present which either allows for a conversion process which
must be first exothermic and then endothermic or else the heat
conversion must allow for very efficient heat transfers. Although, it
has been suggested that the conversion of neutron matter into strange
matter is either direct, through two-flavor quark matter $[3]$, or
through a quantum tunnelling process $[7]$. However, it is unlikely
that the conversion has involved reactions which are initially
exothermic but endothermic in their final stages. Next, it
has been theorized that the core of a neutron state may contain a
superfluid state, which permits very efficient heat transfers. It is
not clear whether or not the superfluid state may exist near the burn
front. Because $v_n$ is typically close to the speed of light, it may
be faster than the critical velocity of the superfluid.

There is one other mechanism which may allow for very effcient heat
transfers which was first suggested by Horvath and Benvenuto $[6]$.
Because the Reynolds numbers for both fluids are extremely large,
it may be that the heat transfer mechanism is not due to thermal
conduction but rather by turbulent convection of the two materials.
This will allow for much more efficient heat transfers. Unlike
Horvath and Benevenuto who argued that the burning will quickly reach
detonation, we find that because $v_q$ is always greater than
the speed of sound in the strange matter, detonation will never
occur.

Most of the parameter space which would allow for the burning of the
star is taken up by the absolutely unstable burning region. In this
region of the parameter space where (condition for absolute unstable
burning) the burn front is absolutely unstable, meaning that any
perturbation of the planar burn front will very quickly grow and
there will be a turbulent mixing of the nuclear and strange matter.
Modeling the burn front as an infinite flat plane is no longer
realistic and other methods need to be used to analyze the burning.

In conclusion, we find that the conversion of the neutron star into a
strange star will never be due to slow combustion or detonation and
in fact may not even be describable using a simple infinite planar
model of the burn front. If it happens at all, then the conversion
could be a fast combustion without detonation. The basic culprit is
$v_q$ which we have found to be greater than the speed of sound in
the strange matter for most of the parameter space. As
such, detonation will never take place; but then the conditions for
stable burning will also be difficult to satisfy. We note, however,
that all models of the burning, including our own, used the infinite
plane description of the burn front. The neutron star is a sphere,
however, and it may be more realistic to use a spherical burn front
to model the burning. As the boundary conditions for a spherical
geometry is quite different from the infinite plane geometry, namely
that the strange matter must be at rest after the burning, a
parameter space which will allow for stable burning may be found. If
not, however, then the conversion of the neutron star will most
probably be absolutely unstable and some way would need to be found
to model the turbulent mixing of the strange and nuclear matter. We
are in the process of analyzing the burning using a spherical geometry.

In more realistic situations where $m_s$, $\alpha_s$ and $Y_e$ are
not zero, thermal energy may be lost by emission of neutrinos
produced in URCA processes and the assumption of energy conservation
may no longer hold. We note, however, that the neutrino emissivity
$e_\nu$ under URCA processes $[11]$: $e_\nu\cong 4.6\times 10^{32}
\left(T\over {\rm MeV}\right)^6 {\rm erg}\>{\rm cm}^{-3}{\rm
sec}^{-1}$ where we have taken $\alpha_s=0.5$, $Y_e=0.01$,
$n_q=3.4\times 10^{38}$ cm$^{-3}$. The heat production rate per unit
volume $e_s\equiv(e_q-e_q(T=0))/t$ for the burning can be calculated
from Eq. $(2)$. For $T\le$10MeV and $\mu \cong (\pi^2 n_q)^{1\over 3}
\cong$ 295MeV, we find that $e_s/e_q \cong 0.18 ({\rm Mev}/T)^4({\rm
sec}/t)$. From the above, we expect the burn front to travel near the
speed of light for $T\le 10$MeV. Taking the star radius as $10$km and
$v_n\cong 0.1$, we find that $t\approx R/v_n\approx 3.3\times
10^{-4}$sec. Our analysis may be used as long as
$\epsilon_s/\epsilon_\nu\gg 1$, so that $T<4.8$MeV. As we know $T$
may reach $10$MeV or higher. Note, however, that for $T=10$MeV, the
neutrino mean free path $[11]$, $l_\nu \approx 3$m$<<R$.
Consequently, our method of analyzing the burning process will be
applicable even in the more realistic model because neutrino emission
is relatively small at low temperatures and at high temperatures,
neutrino trapping becomes efficient.

We have also analysed a case in which the equation of state of the neutron gas
is of BJ type and $T=10$ MeV. The result is quite similar to Fig. 3.
Moreover, we could not find any choices of $B$ and $n_n$ for $T=10$ MeV and
$m_s=0$ which would reproduce Olinto's result.

Preliminary results for the case of finite strange quark mass have shown no
qualitative changes to the parameter space. Non-zero strong coupling
corrections should have similar results.

\vskip1truecm
\centerline{\bf Acknowledgements}
This work is supported by the National Science Council of the Republic of
China under contract numbers NSC 82-0208-M-001-086 and NSC
82-0208-M-001-131-T.
\vfill
\supereject
\centerline{\bf REFERENCES}
\item{$[1]$}G. Baym, E. W. Kolb, L. McLerran, and T. P. Walker, Phys. Lett.
                {\bf 160B}, 181 (1985).
\par
\item{$[2]$}A. V. Olinto, Phys. Lett. {\bf 192B}, 71 (1987).
\par
\item{$[3]$}A. V. Olinto, Nucl. Phys. B (Proc. Suppl.) {\bf B24}, 103 (1991).
\par
\item{$[4]$}M. L. Olesen and J. Madsen, Nucl. Phys. B (Proc. Suppl.)
                {\bf B24}, 170 (1991).
\par
\item{$[5]$}H. Heiselberg, G. Baym, and C. J. Pethick, Nucl. Phys. B
                (Proc. Suppl.) {\bf B24}, 144 (1991).
\par
\item{$[6]$}J. E. Horvath and O. G. Benvenuto, Phys. Lett. {\bf 213B}, 516
                (1988).
\par
\item{$[7]$}O. G. Benvenuto, J. E. Horvath, and H. Vucetich, Nucl.
                Phys. B (Proc. Suppl.) {\bf B24}, 125 (1991).
\par
\item{$[8]$}E. Witten, Phys. Rev. {\bf D30}, 272 (1984).
\par
\item{$[9]$}E. Fahri and R. L. Jaffe, Phys. Rev. {\bf D30} 2379 (1984).
\par
\item{$[10]$}L. D. Landau and E. M. Lifshitz, {\sl Fluid Mechanics},
                Chapters 14 and 15 (Pergamon Press, New York 1959).
\par
\item{$[11]$}N. Iwamoto, Ann. Phys. {\bf 141} 1 (1982).
\par
\item{$[12]$}S. L. Shapiro and S. A. Teukolsky, {\sl Black Holes, White
                Dwarfs, and Neutron Stars}, Chapter 8 (John Wiley \& Sons,
                New York, 1983).
\par
\item{$[13]$}L. D. Landau and E. M. Lifshitz, {\sl Statistical Physics,
                Part 1, 3rd ed.}, Chapter 5 (Pergamon Press, New York 1980).
\vfill
\end